\documentclass[12pt]{ORSNZ}
\usepackage{amssymb, vmargin, ORSNZ}
\usepackage{optidef}
\usepackage{glossaries}
\usepackage{microtype}
\usepackage{booktabs}
\usepackage{multirow}
\newacronym{SCVX}{SCVX}{successive convex optimisation}
\newacronym{SCP}{SCP}{sequential convex programming}
\newacronym{FOH}{FOH}{first-order-hold discretisation}
\newacronym{LLO}{LLO}{low-altitude lunar orbit}
\newacronym{eLLO}{eLLO}{extremely Low-Lunar Orbit}
\newacronym{ZOH}{ZOH}{zero-order-hold}
\newacronym{LRO}{LRO}{Lunar Reconnaissance Orbiter}
\newacronym{ICRF}{ICRF}{international celestial reference frame}
\newacronym{DCM}{DCM}{direction cosine matrix}
\newacronym{LME2000}{LME2000}{lunar mean equator of date J2000}
\newacronym{GRAIL}{GRAIL}{Gravity Recovery And Interior Laboratory}
\newacronym{SDF}{SDF}{Signed Distance Field}
\newacronym{AD}{AD}{automatic differentiation}
\newacronym{SRP}{SRP}{solar radiation pressure}
\newacronym{DDP}{DDP}{differential dynamic programming}
\newacronym{SEP}{SEP}{solar electric propulsion}

\setpapersize{A4}
\setmargnohfrb{30mm}{20mm}{30mm}{20mm} 
\title{Sequential Convex Programming for Multimode Spacecraft Trajectory Optimization}
\author{J. Yarndley\\Te P\=unaha \=Atea -- Space Institute\\University of
  Auckland\\New Zealand\\jyar540\@@aucklanduni.ac.nz} 

\date{} 
\begin{document}

\maketitle
\pagestyle{empty} \thispagestyle{empty}
\begin{abstract}
Spacecraft equipped with multiple propulsion modes or systems can offer enhanced performance and mission flexibility compared with traditional configurations. Despite these benefits, the trajectory optimization of spacecraft utilizing such configurations remains a complex challenge. This paper presents a sequential convex programming (SCP) approach for the optimal design of multi-mode and multi-propulsion spacecraft trajectories. The method extends the dynamical linearization within SCP using sparse automatic differentiation, enabling efficient inclusion of multiple propulsion modes or systems without complex manual reformulation while maintaining comparable computational efficiency. New constraint formulations are introduced to ensure selection of a single propulsion mode at each time step and limit the total number of modes used. The approach is demonstrated for (i) a low-thrust Earth-67P rendezvous using the SPT-140 thruster with 20 discrete modes, and (ii) an Earth-Mars transfer employing both a low-thrust engine and a solar sail. Results confirm that the proposed method can efficiently compute optimal trajectories for these scenarios.

\textbf{Key words:} convex programming, trajectory optimization, optimal control, mission design, solar electric propulsion, solar sails
\end{abstract}

\section{Introduction}

Recent advances in low-thrust technologies for spacecraft propulsion, particularly in \gls{SEP}, have revolutionized the design and capability of modern space missions. With significantly higher propellant efficiency, these systems provide clear advantages over conventional chemical propulsion, but this comes at the cost of lower thrust levels and high continuous power demands. The demonstrated success of \gls{SEP} in missions such as \textit{Dawn} \cite{russellDawnMissionMinor2012} and \textit{Psyche} \cite{ohDevelopmentPsycheMission2019} highlights the maturity, versatility and applicability of these technologies. Furthermore, the development of propellantless propulsion concepts, such as solar sailing, have also opened further avenues for mission design. However, these tend to exhibit even lower thrust levels and control constraints when compared to \gls{SEP} \cite{gongReviewSolarSail2019}.

Despite these advances, the design of optimal low-thrust trajectories remains a complex challenge. This complexity is compounded when considering that \gls{SEP} engines often operate in multiple discrete modes, each with different thrust levels and efficiencies \cite{roveyReviewMultimodeSpace2020}. For example, the SPT-140 \gls{SEP} engine have been analyzed with 20 discrete operation modes with thrusts ranging from 87 to 287 mN and specific impulses from 925 to 1929 s \cite{manzellaPerformanceEvaluationSPT1401997}. Consequently, inefficient approximation and selection of the optimal operation mode at each time step can have significant impacts on the overall mission performance.

In terms of propellantless propulsion concepts, such as solar sails, the limitations from the lens of trajectory design are even more pronounced. Their underlying physics induces a nonlinear coupling between the achievable acceleration directions and magnitudes \cite{mcinnesSolarSailing1999}, and more fundamentally, the maximum acceleration from a solar sail is often many orders of magnitude lower than chemical and \gls{SEP} systems \cite{gongReviewSolarSail2019,spencerSolarSailingTechnology2019}.

Therefore, care must be taken in the selection of appropriate optimization techniques. Amongst a wide range of spacecraft trajectory optimization techniques \cite{chaiReviewOptimizationTechniques2019}, such as those based on direct methods, indirect methods and \gls{DDP}, there has recently been a growing interest in \gls{SCP}. \gls{SCP} is a direct method based on convex programming that iteratively solves a sequence of convex subproblems that approximate the original non-convex problem \cite{maoSuccessiveConvexificationSuperlinearly2019,malyutaConvexOptimizationTrajectory2022}. The main advantages of \gls{SCP} lies in the efficiency, robustness, and convergence guarantees offered by convex programming. \gls{SCP}-based methods have been widely used across the aerospace domain, including in space missions with conventional and low-thrust engines \cite{hofmannComputationalGuidanceLowThrust2023}. Furthermore, even through applications to solar sail trajectory design \cite{songSolarsailDeepSpace2019} have been limited by the difficulty of convexifying the solar sail control, recent work on the lossless convexification of the sail dynamics \cite{oguriLosslessControlConvexFormulation2024} has enabled tractable implementations of solar sail trajectory design problems within \gls{SCP} frameworks.

Current state-of-the-art in the optimal design of multi-mode spacecraft trajectories has primarily focused on indirect methods \cite{roveyReviewMultimodeSpace2020,aryaLowThrustGravityAssistTrajectory2021,taheriNovelApproachOptimal2020,clineIndirectOptimalControl2024}, which find the optimal costates of the system and use these to determine the optimal control at each time step. However, indirect methods are often difficult to implement and require a good initial guess to converge, and there is substantial complexity involved in the introduction of additional constraints or differing propulsion methods \cite{chaiReviewOptimizationTechniques2019}.

To address the challenge of the efficient design of multi-mode and multi-propulsion spacecraft trajectories, this paper presents a \gls{SCP}-based framework that efficiently incorporates multiple propulsion modes and systems and avoids manual reformulation of the optimal control problem. The key dynamical linearization within \gls{SCP} is extended using sparse automatic differentiation to efficiently include multiple propulsion modes. New lossless constraint formulations are introduced to ensure selection of a single propulsion mode at each time step and limit the total number of modes used. The proposed methodology is demonstrated on (i) a low-thrust Earth-67P rendezvous using the SPT-140 thruster with 20 discrete modes and (ii) an Earth-Mars transfer employing both a low-thrust engine and a solar sail.

\section{Methodology}

In this section the dynamical environment is presented along with the control transcription and optimization methodology used throughout this work. Firstly, the two-body dynamics, \gls{SEP} and solar sail propulsion models are introduced and then the \gls{SCP} framework is presented, including the convexification of the dynamics and the full problem formulation.

\subsection{Dynamics and propulsion models}

The state vector of the spacecraft is defined in Cartesian coordinates as $\boldsymbol{x} = [\boldsymbol{r}, \boldsymbol{v}, m]^T$, where $\boldsymbol{r}$ and $\boldsymbol{v}$ are the position and velocity vectors of the spacecraft and $m$ is the spacecraft mass. The dynamics of the spacecraft are determined by two-body gravitational acceleration between the spacecraft and the Sun only, expressed as the standard Newtonian point-mass gravity. Additionally, several control terms are added to the dynamics for each of the low-thrust and solar sail propulsion models. The dynamics are expressed in the Cartesian frame as follows,
\begin{equation}
\dot{\boldsymbol{x}} = \begin{bmatrix}\dot{\boldsymbol{r}}(t) \\ \dot{\boldsymbol{v}}(t) \\ \dot{m}(t) \end{bmatrix} = \begin{bmatrix}\boldsymbol{v} \\ -\frac{\mu_\odot}{r^3} \boldsymbol{r} + \boldsymbol{a}_{\text{SEP}} + \boldsymbol{a}_{\text{SAIL}} \\ \dot{m}_{\text{SEP}} \end{bmatrix},
\end{equation}
where $\boldsymbol{a}_{\text{SEP}}$ is the acceleration due to \gls{SEP} propulsion, $\boldsymbol{a}_{\text{SAIL}}$ is the acceleration due to solar sailing, $\dot{m}_{\text{SEP}}$ is the mass flow rate of the \gls{SEP} propulsion, $\mu_\odot$ is the gravitational parameter of the Sun, and $r = \|\boldsymbol{r}\|$ is the distance of the spacecraft from the Sun.

The \gls{SEP} propulsion model assumes a control vector $\boldsymbol{T}_i$ with $\| \boldsymbol{T}_i \| \leq 1$ which defines the thrust direction and throttle factor for each thrust mode $i$. This can change over time depending on the thrust transcription strategy; here we assume it is held constant throughout each segment of the trajectory as in a \gls{ZOH} transcription. The thrust acceleration and mass flow rate for a set of \gls{SEP} modes can then be expressed as,
\begin{align}
    \boldsymbol{a}_{\text{SEP}} &= \sum_{i=1}^{N} \frac{T_{\text{max}, i}}{m_i} \boldsymbol{T}_i, \\
    \dot{m}_{\text{SEP}} &= \sum_{i=1}^{N} -\frac{\|\boldsymbol{T}_i\|}{I_{\text{sp},i} g_0},
\end{align}
where $I_{\text{sp}}$ is the specific impulse of the thruster and $g_0 = 9.80665$ m/s$^2$ is the standard gravity. The maximum thrust $T_{\text{max}, i}$ and specific impulse $I_{\text{sp},i}$ of each thruster $i$ can be constant or vary with time, for example as a function of distance from the Sun. Power limitations for \gls{SEP} operation modes can also be included in this manner.  

The non-ideal flat plate solar model is used to model the \gls{SRP} acceleration of the solar sail. A detailed description of this model is presented by \cite{oguriSolarSailingPrimer2022}. This model expresses the total \gls{SRP} acceleration acting on the sail as the sum of specular reflection, diffuse reflection and absorption components, which either act along the normal or sunlight directions. The total \gls{SRP} acceleration, parameterised by the sail normal direction $\hat{\boldsymbol{u}}_n$, is therefore,
\begin{equation}\label{srp_flat_plate_acceleration}
\boldsymbol{a}_{\text{SAIL}} = -\frac{CA}{m}\left(\frac{r_\oplus}{r}\right)^2 \left( \hat{\boldsymbol{u}}_n \cdot \hat{\boldsymbol{u}}_r \right) 
\left[ \underbrace{2\nu \,\hat{\boldsymbol{u}}_n}_{\text{diffuse}} + \underbrace{4\mu \left(\hat{\boldsymbol{u}}_n \cdot \hat{\boldsymbol{u}}_r  \right) \hat{\boldsymbol{u}}_n}_{\text{specular}} + \underbrace{\left(1 - 2\mu\right) \hat{\boldsymbol{u}}_r}_{\text{absorption}} \right],
\end{equation}
where $r_\oplus$ is the reference distance from the sun (typically 1 AU), $C$ is the solar flux at $r_\oplus$, $A$ is the sail area, $m$ is the spacecraft mass, $r$ is the distance of the spacecraft from the Sun, $\hat{\boldsymbol{u}}_n$ is the unit normal vector of the sail, $\hat{\boldsymbol{u}}_r$ is the unit vector from the spacecraft to the Sun, $\mu$ is the solar sail specular reflection coefficient, and $\nu$ is the solar sail diffuse reflection coefficient. In this work, the sail normal $\hat{\boldsymbol{u}}_n$ is parameterized for output in terms of the cone angle $\alpha$ and clock angle $\beta$, which define the orientation of the sail normal with respect to the sunlight direction $\hat{\boldsymbol{u}}_r$. 



The equations of motion are rescaled and non-dimensionalized to improve numerical behavior during integration and optimization. The distance unit is defined as $\mathrm{DU} = 1~\mathrm{AU} = 1.495979\times10^8~\mathrm{km}$. The gravitational parameter of the Sun is set to $\mu_\odot = 1$ DU$^3$/TU$^2 = 1.327124\times10^{11}$ km$^3$/s$^2$, from which the time unit is derived as
\begin{equation*}
  \mathrm{TU} = \sqrt{\frac{\mathrm{DU}^3}{\mu_\odot}} = 5.022643\times10^6~\mathrm{s}.
\end{equation*}
The corresponding velocity unit is $\mathrm{VU} = \mathrm{DU}/\mathrm{TU} = 29.784692~\mathrm{km/s}$, and the mass unit $\mathrm{MU}$ is set to the initial mass of the spacecraft.

\subsection{Sequential convex programming}

Applying the general approach of SCP \cite{malyutaConvexOptimizationTrajectory2022}, the trajectory is split into many parts as in a direct method, each referred to as a segment with index $n= 1, 2, ..., N$. The number of segments $N$ is calculated based on an intended timespan between potential impulses. With this discretization, the linearized dynamic constraints are constructed around a reference trajectory, which requires an appropriate initial guess. For this work, a ballistic reference trajectory with zero controls is used. Then, each segment is assigned a single control input for each propulsion mode, which are held constant throughout the segment as in a \gls{ZOH} transcription. For \gls{SEP} propulsion, this is the thrust direction and throttle factor $\boldsymbol{T}_i$ for each mode $i$. For solar sailing, it is the sail acceleration vector $\boldsymbol{a}_{\text{SAIL}}$.

Then, given the reference trajectory ($\boldsymbol{\bar{x}_n}, \boldsymbol{\bar{T}_{n, i}}, \boldsymbol{\bar{a}_{\text{SAIL}, n}}$), the dynamics, and the segmentation $n = 1, 2, ..., N$, a discrete form of the spacecraft dynamics is obtained,
\begin{align} \label{equation_dynamics_linear}
    \forall n : \boldsymbol{x}_{n+1} = \boldsymbol{A}_n\boldsymbol{x}_n + \sum_{i=1}^N \boldsymbol{B}_{n, i}\boldsymbol{T}_{n, i} + \boldsymbol{C}_{n} \boldsymbol{a}_{\text{SAIL}, n} + \boldsymbol{d}_n + \boldsymbol{e}_n,
\end{align}
where $\boldsymbol{e}_n$ is a virtual control variable used to ensure feasibility of the linearized dynamics. The virtual control is penalized in the objective function to ensure it is only used when necessary, and is constrained to be positive. The remaining terms are defined as,
\begin{align}
    \boldsymbol{A}_n &= \left. \left[\frac{\partial}{\partial \boldsymbol{x}} \int_{t_n}^{t_{n+1}}\dot{\boldsymbol{x}}\,\text{d}t \right]\right|_{(\boldsymbol{\bar{x}_n}, \boldsymbol{\bar{T}_{n, i}}, \boldsymbol{\bar{a}_{\text{SAIL}, n}})}\\
    \boldsymbol{B}_{n, i} &= \left. \left[\frac{\partial}{\partial \boldsymbol{T}_{n, i}} \int_{t_n}^{t_{n+1}} \dot{\boldsymbol{x}}\,\text{d}t \right]\right|_{(\boldsymbol{\bar{x}_n}, \boldsymbol{\bar{T}_{n, i}}, \boldsymbol{\bar{a}_{\text{SAIL}, n}})} \\ 
    \boldsymbol{C}_n &= \left. \left[\frac{\partial}{\partial \boldsymbol{a}_{\text{SAIL}, n}} \int_{t_n}^{t_{n+1}} \dot{\boldsymbol{x}}\,\text{d}t \right]\right|_{(\boldsymbol{\bar{x}_n}, \boldsymbol{\bar{T}_{n, i}}, \boldsymbol{\bar{a}_{\text{SAIL}, n}})} \\ 
    \boldsymbol{d}_n &= \boldsymbol{\bar{x}}_{n+1} - \boldsymbol{A}_n \boldsymbol{\bar{x}_n} - \sum_{i=1}^N \boldsymbol{B}_{n, i} \boldsymbol{T}_{n, i} - \boldsymbol{C}_n \boldsymbol{a}_{\text{SAIL}, n}.
\end{align}

The matrix $\boldsymbol{A}_n$ is the state transition matrix (STM) representing the changes in the final state $\boldsymbol{x}_{n+1}$ of each segment with respect to the initial state $\boldsymbol{x}_{n}$. Correspondingly, $\boldsymbol{B}_{n, i}$ represents the changes in the final state of each segment with respect to the \gls{SEP} mode $i$ and $\boldsymbol{C}_n$ represents the changes in the final state of each segment with respect to the solar sail acceleration. The vector $\boldsymbol{d}_n$ is a constant term that ensures that the linearization is exact at the reference trajectory.

Rather than using an analytic formulation, the partial derivatives are computed via \gls{AD} which is directly applied to the initial conditions of a numerical integration solver. The \texttt{Tsit5} numerical integrator is used from the \texttt{DifferentialEquations.jl} \cite{rackauckasDifferentialEquationsJlPerformant2017} library with absolute tolerance $10^{-10}$ and relative tolerance $10^{-10}$. The \gls{AD} is computed in forward mode through the use of \texttt{ForwardDiff.jl} \cite{revelsForwardModeAutomaticDifferentiation2016}.

\begin{figure}[h]
\centering
\includegraphics[width=\textwidth]{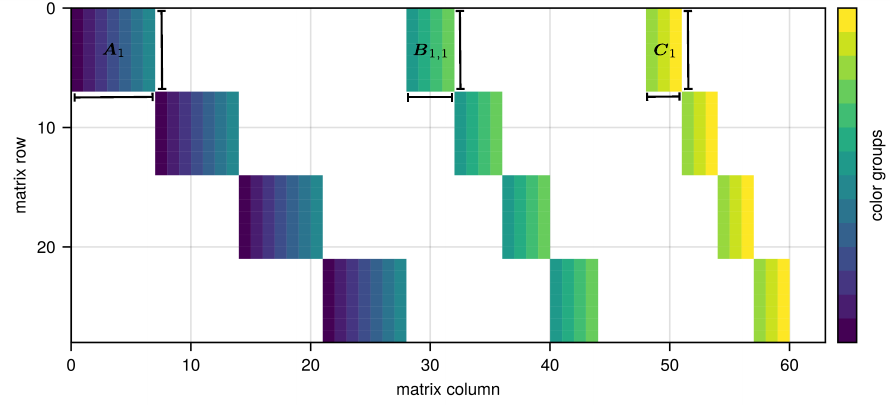}
\caption{Structure and colors assigned to the Jacobian matrix.}
\label{figure_matrix_coloring}
\end{figure}

Sparse \gls{AD} \cite{hillSparserBetterFaster2025} is employed to improve the computational efficiency of computing the entire dynamics constraint matrix, improve flexibility and to automatically detect the non-zero entries. Instead of computing each $\boldsymbol{A}$, $\boldsymbol{B}_i$ and $\boldsymbol{C}$ matrix for each segment, the entire dynamics constraint matrix can be computed efficiently in the minimal number of calls through matrix coloring and subsequent rearrangement. This matrix rearrangement process allows for the efficient use of forward-mode \gls{AD} which can compute columns in a single pass. This avoids the need to manually compute and design a discretization scheme for the dynamical constraint, enabling a significant improvement in flexibility and ease of implementation of various dynamical models. The resulting constraint matrix and coloring pattern is illustrated in Fig.~\ref{figure_matrix_coloring} for a problem with $N=5$ segments, $1$ \gls{SEP} mode and $1$ solar sail. The sparsity pattern is clearly visible, and the coloring pattern shows that only $14$ calls of forward-mode \gls{AD} (one for each color) are required to compute the entire constraint matrix, rather than the 63 calls (one for each column) without coloring.


Hard trust region constraints are introduced to ensure that the linearization of the dynamical constraint remains accurate. The trust region is selected to have a constant size that does not change as the \gls{SCP} algorithm progresses for simplicity. This constraint takes the form
\begin{align} \label{equation_state_trust_regions}
    \forall {n}: -\boldsymbol{\epsilon}_1 \leq \boldsymbol{x}_n -\bar{\boldsymbol{x}}_n \leq \boldsymbol{\epsilon}_1.
\end{align}
Several values for the initial size of the trust regions were tested, $\boldsymbol{\epsilon}_1 = 10^{-1}$ is selected for this work. Next, to obtain the norms of the \gls{SEP} controls and constraint the maximum throttle, a second-order-cone (SOC) constraint is used,
\begin{equation} \label{equation_sep_control}
\begin{aligned} 
    \forall n, i &: \|\boldsymbol{T}_{n, i}\|_2 \leq T_{n, i}, \\
    \forall n, i &: T_{n, i} \leq 1. 
\end{aligned}
\end{equation}
This ensures that the norm of the control vector can be used to calculate the mass flow rate in the dynamical constraints, and that the throttle factor of each \gls{SEP} mode is between 0 and 1. The SOC constraint is lossless and is binding at optimality if an objective that implicitely minimizes the use of \gls{SEP} thrust is used. Two additional constraints are introduced to ensure that only one \gls{SEP} mode is active at each time step, and that the total number of modes used throughout the trajectory is limited. Firstly, to ensure only one mode is active at each time step, the following constraint is introduced,
\begin{align}
    \forall n &: \sum_{i=1}^N T_{n, i} \leq 1. \label{equation_sep_one_mode}
\end{align}
Because the marginal costs of the \gls{SEP} modes are different, this constraint will always ensure only a single mode is active at each time step. Only in cases where the mode switch happens within the segment could a second mode also be active, but their total throttle will still be limited to 1. Secondly, to limit the total number of modes used throughout the trajectory, the following set of constraints is introduced,
\begin{equation} \label{equation_sep_max_modes_all}
\begin{aligned}
    \forall i, n &: T_{n, i} \leq k_{i}, \\ 
    \forall i &: b_i \in \{0, 1\}, \\
    \forall i &: k_i \leq 10^6\, b_i, \\
    &\sum_{i=1}^N b_i \leq K.
\end{aligned}
\end{equation}
where $k_i$ is an auxiliary variable for each mode $i$ representing the maximum throttle of that mode throughout the entire trajectory, $b_i$ is a binary variable, and $K$ is the maximum number of modes allowed. This constraint changes the convex subproblem to require a mixed-integer convex programming solver, but is relatively lightweight due to the small number of binary variables required. 

The solar sail acceleration can be constrained to be within the physically achievable limits through a lossless convexification procedure which is presented by \cite{oguriLosslessControlConvexFormulation2024} which is not repeated here for brevity. The procedure introduces further SOC constraints and also performs a second-order linearization on the solar sail acceleration envelope.

The initial and final state constraints are simply
\begin{equation} \label{equation_boundary_states}
\begin{aligned}
    \boldsymbol{x}_{1} &= \boldsymbol{x}_{\text{initial}}, \\
    \boldsymbol{x}_{N} &= \boldsymbol{x}_{\text{final}}.
\end{aligned}
\end{equation}

with $\boldsymbol{x_{\text{initial}}}$ the initial state and $\boldsymbol{x_{\text{final}}}$ the final state. Then the main objective is to maximize the final mass whilst minimizing the use of virtual controls. This is achieved through the following objective function,
\begin{align} \label{equation_scp_objective}
    J = -m_N + 10^2 \sum_{n=1}^N \|\boldsymbol{e}_n\|_1,
\end{align}
The choice of $10^2$ for penalization tended to work well in our testing, and the 1-norm is computed as the sum of the absolute values of the elements of $\boldsymbol{e}_n$ through a linear formulation. Therefore, the entire optimization problem for all cases is
\begin{mini*}
    {}{\eqref{equation_scp_objective}\quad\text{(objective function $J$)}}{}{}
    \addConstraint{\eqref{equation_dynamics_linear}}{}{\quad\text{(linearized dynamics)}}
    \addConstraint{\eqref{equation_state_trust_regions}}{}{\quad\text{(state hard trust regions)}}
    \addConstraint{\eqref{equation_sep_control}}{}{\quad\text{(control magnitude limits)}}
    \addConstraint{\eqref{equation_sep_one_mode}}{}{\quad\text{(one SEP mode per time step)}}
    \addConstraint{\eqref{equation_sep_max_modes_all}}{}{\quad\text{(max SEP modes used)}}
    \addConstraint{\eqref{equation_boundary_states}}{}{\quad\text{(initial and final states)}}
    \addConstraint{}{}{\quad\text{(solar sail acceleration)}}
    \label{equation_full_scp_problem}.
\end{mini*}
The \gls{SCP} process repeatedly solves this convex problem and updates the linearized constraint \eqref{equation_dynamics_linear} and the constraints relating to the convexification of the solar sail control with the optimal solution from the previous iteration. The convergence of the algorithm is determined by the accuracy of the linearization compared to the truth from numerical propagation. In terms of implementation, \texttt{JuMP.jl} \cite{lubinJuMPRecentImprovements2023} is used to create and modify the convex problems, and MOSEK \cite{mosekapsMOSEKOptimizerAPI2025} is used to solve them. 

\section{Results and Discussion}

This section details the application of the proposed \gls{SCP} framework to two example problems, an Earth-67P rendezvous and an Earth-Mars transfer.

\subsection{Earth-67P rendezvous with SPT-140 thruster}

A rendezvous mission from Earth to comet 67P/Churyumov-Gerasimenko is considered using the SPT-140 \gls{SEP} engine with 20 discrete operation modes \cite{aryaLowThrustGravityAssistTrajectory2021}. Throughout the trajectory the maximum thrust and specific impulse of each mode are held constant. The spacecraft is powered using a solar power system which provides $10~\text{kW}$ at $1~\text{AU}$, which decreases with the inverse square of the distance from the Sun. If a mode requires more power than is available, the thruster is forced to produce no thrust even if the mode is selected.


The mission has a fixed time-of-flight of 1776 days and an initial spacecraft mass of $m_\text{initial}=2500$ kg. Initial and final states are provided as
\begin{align*}
    \boldsymbol{r}_\text{initial} &= [-1671985.956644, -151914424.309981, 1699.375105]^\mathsf{T} \text{ km}, \\
    \boldsymbol{v}_\text{initial} &= [29.307044, -0.596900, -0.000411]^\mathsf{T} \text{ km/s}, \\
    \boldsymbol{r}_\text{final} &= [-465627493.144610, -50530561.307303, 40190127.950002]^\mathsf{T} \text{ km}, \\
    \boldsymbol{v}_\text{final} &= [-9.721779, -14.629481, -0.234945]^\mathsf{T} \text{ km/s}.
\end{align*}

State and control discretization is performed using an intended segment timespan of 5 days, resulting in a total of $N=356$ segments. The initial guess is a trajectory interpolated across the modified equinoctial orbital elements \cite{walkerSetModifiedEquinoctial1985} of the initial and final states with zero thrust for all modes. 

\begin{figure}[h]
\centering
\includegraphics[width=\textwidth]{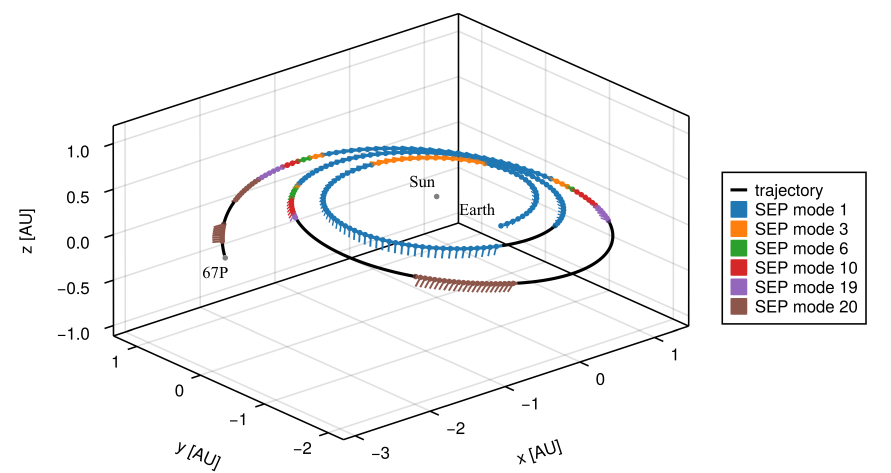}
\caption{Optimal Earth-67P trajectory with all 20 modes of the SPT-140 thruster.}
\label{figure_67P_trajectory}
\end{figure}

\begin{figure}[h]
\centering
\includegraphics[width=\textwidth]{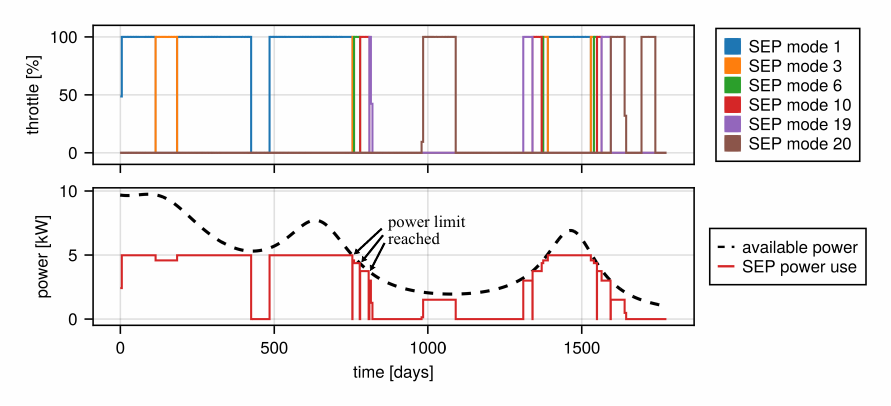}
\caption{Control and power profiles for the Earth-67P mission.}
\label{figure_67P_controls}
\end{figure}

The optimal trajectory found using the proposed \gls{SCP} framework with a maximum of 20 modes is illustrated in Fig.~\ref{figure_67P_trajectory}, where the spacecraft trajectory and the thrusting directions are shown in the inertial frame. A total of 6 modes are used throughout the trajectory, with a final mass of $m_\text{final}=1157.659$ kg which corresponds to a total propellant consumption of $1342.341$ kg. Correspondingly, the control profile and activation of the power limitation constraint is shown in Fig.~\ref{figure_67P_controls}. It is apparent that the power limitation is adhered to throughout the trajectory. Intrestingly, there are some very small periods where the power limitation comes into effect within a segment but the thruster is still active. This indicates that even through the power limitation is violated at some times throughout the segment, it is still optimal to thrust (while possible) rather than selecting a different operation mode.

\begin{table}[ht!]
\centering
\scriptsize
\setlength{\tabcolsep}{6pt}
\renewcommand{\arraystretch}{1.05}
\caption{Analysis of selected SPT-140 modes for Earth--67P transfer.}
\begin{tabular}{l r r r r r}
\toprule
\textbf{Max} & \textbf{Total propellant}  & \textbf{Active} & \textbf{Startups} & \textbf{Burn time} & \textbf{Propellant} \\
\textbf{modes} & \textbf{use [kg]} & \textbf{modes} & \textbf{[\#]} & \textbf{[days]} & \textbf{use [kg]} \\
\midrule
\multirow[t]{2}{*}{2} & \multirow[t]{2}{*}{1395.589} & 3  & 4 & 825.441 & 1267.264 \\
                      &                             & 19 & 3 & 129.935 & 128.325 \\
\midrule
\multirow[t]{3}{*}{3} & \multirow[t]{3}{*}{1366.703} & 1  & 4 & 720.876 & 865.920 \\
                      &                             & 3  & 5 & 196.269 & 301.330 \\
                      &                             & 19 & 3 & 201.957 & 199.453 \\
\midrule
\multirow[t]{4}{*}{4} & \multirow[t]{4}{*}{1351.266} & 1  & 3 & 843.339 & 1013.022 \\
                      &                             & 6  & 3 & 72.125  &  87.332  \\
                      &                             & 19 & 3 & 162.996 & 160.976  \\
                      &                             & 20 & 2 & 170.010 &  89.936  \\
\midrule
\multirow[t]{5}{*}{5} & \multirow[t]{5}{*}{1344.314} & 1  & 3 & 834.404 & 1002.289 \\
                      &                             & 6  & 3 &  68.985 &   83.529 \\
                      &                             & 10 & 3 &  70.000 &   84.291 \\
                      &                             & 19 & 3 &  81.467 &   80.460 \\
                      &                             & 20 & 2 & 177.217 &   93.746 \\
\midrule
\multirow[t]{6}{*}{20} & \multirow[t]{6}{*}{1342.341} & 1  & 4 & 762.566 & 915.996 \\
                        &                              & 3  & 4 & 100.000 & 153.529 \\
                        &                              & 6  & 3 & 37.107  & 44.927  \\
                        &                              & 10 & 3 & 67.145  & 80.852  \\
                        &                              & 19 & 3 & 66.813  & 65.984  \\
                        &                              & 20 & 2 & 153.223 & 81.053  \\
\bottomrule
\end{tabular}
\label{tab:67P_modes_optimal}
\end{table}

An additional analysis was conducted to determine the effect of limiting the maximum number of modes used throughout the trajectory on the final mass. The results are summarized in Table~\ref{tab:67P_modes_optimal}. Since only 6 modes are optimally selected in the unconstrained case, only additional restrictions of 5 modes or less are considered. As would be expected, with less available modes, the propellant use increases and correspondingly the final mass decreases. With a single mode, the problem becomes infeasible as power limitations prevent the use of the higher thrust modes when far from the Sun and lower thrust modes cannot provide sufficient total acceleration throughout the trajectory. Interestingly, the selection of optimal modes is not always a subset of the modes selected with higher limits. For example, with a 3 mode limit, mode 3 is selected, which is not present in the 4 and 5 mode limited cases.

\subsection{Earth-Mars transfer with two different propulsion technologies}

Next, an Earth-Mars transfer mission is considered using both a \gls{SEP} engine and a solar sail. The spacecraft parameters for this example are based on the \textit{NEA Scout} mission \cite{lantoineTrajectoryManeuverDesign2024} but additionally with a small \gls{SEP} engine. The mission has a fixed time-of-flight of 1600 days and an initial spacecraft mass of $m_\text{initial}=11.629$ kg. The spacecraft is equipped with a solar sail of area $84.6$ m$^2$ with $\mu=0.40495$ and $\nu=0.014957$, and the reference solar flux at 1 AU is $4.5391~\mathrm{\mu N/m^2}$. The \gls{SEP} engine has a maximum thrust of $500~\mathrm{\mu N}$ and a specific impulse of $1000$ s. Initial and final states are derived from the true positions of Earth and Mars assuming a mission start time of 2027-08-01T00:00:00.0 UTC, and are given as
\begin{align*}
    \boldsymbol{r}_\text{initial} &= [93853872.843842, -119373984.588205, 7038.663554]^\mathsf{T} \text{ km}, \\
    \boldsymbol{v}_\text{initial} &= [22.932995, 18.299722, -0.001290]^\mathsf{T} \text{ km/s}, \\
    \boldsymbol{r}_\text{final} &= [207206767.959246, -13772583.781563, -5367172.552103]^\mathsf{T} \text{ km}, \\
    \boldsymbol{v}_\text{final} &= [2.533468, 26.245217, 0.487944]^\mathsf{T} \text{ km/s}.
\end{align*}

\begin{figure}[h]
\centering
\includegraphics[width=\textwidth]{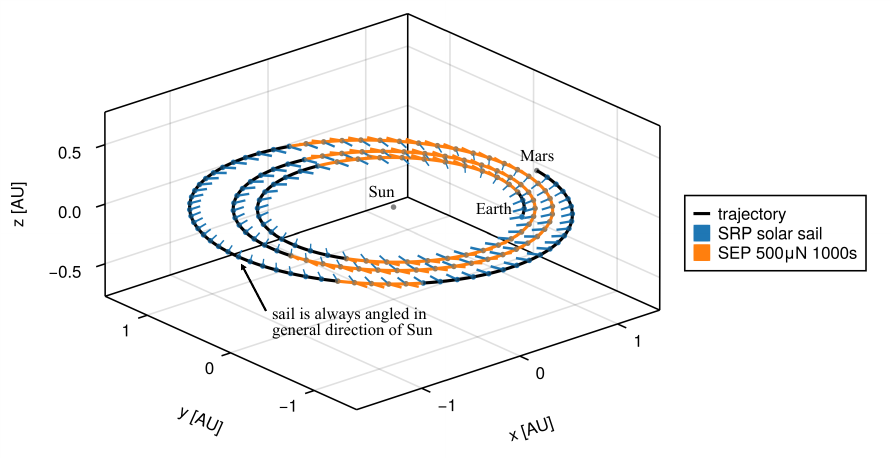}
\caption{Optimal Earth-Mars trajectory with SEP and solar sail.}
\label{figure_mars_trajectory}
\end{figure}

\begin{figure}[h]
\centering
\includegraphics[width=3.25in]{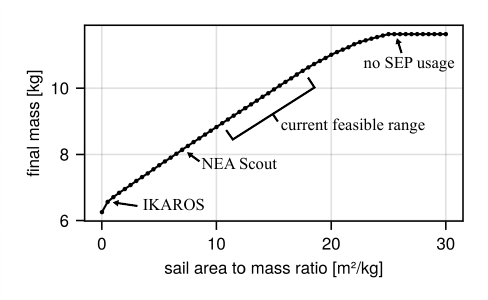}
\caption{Final mass of Earth-Mars mission with differing sail area-to-mass ratios.}
\label{figure_67P_solar_sail_area}
\end{figure}

The optimal trajectory found is illustrated in Fig.~\ref{figure_mars_trajectory}, with a final spacecraft mass of $m_\text{final}=8.198~\text{kg}$. Subsequently, a study was conducted to determine the effect of the sail area-to-mass ratio, which determines the characteristic acceleration from the sail. These results are illustrated in Fig.~\ref{figure_67P_solar_sail_area}, which demonstrate that trajectories which do not use the \gls{SEP} engine at all are possible with area-to-mass ratios greater than $25~\text{m}^2/\text{kg}$. A linear decrease in the final mass is then observed as the area-to-mass ratio decreases. Accordingly, for this mission, current feasible constructions of solar sails could save up to half of the total propellant compared to using \gls{SEP} only.

\section{Conclusions}

This paper presents a \gls{SCP}-based framework for the optimal design of multi-mode and multi-propulsion spacecraft trajectories and demonstrates its application with two example problems. Results confirm that the proposed method can efficiently compute optimal trajectories for these scenarios. The key contributions of this work include the extension of the dynamical linearization within \gls{SCP} using sparse automatic differentiation, enabling efficient solving of multi-mode and multi-propulsion problems without complex manual reformulation. Several new constraint formulations are also introduced to ensure selection of a single propulsion mode at each time step and limit the total number of modes used.

\section*{Acknowledgments}

The author would like to thank Roberto Armellin from Te P\=unaha \=Atea -- Space Institute, University of Auckland and Gregory Lantoine from Jet Propulsion Laboratory, California Institute of Technology for their helpful discussions and suggestions.

\bibliographystyle{achicago}
\bibliography{ref}

@inproceedings{ohDevelopmentPsycheMission2019,
  title = {Development of the {{Psyche Mission}} for {{NASA}}'s {{Discovery}}  {{Program}}},
  booktitle = {36th {{International Electric Propulsion Conference}}},
  author = {Oh, David Y. and Collins, Steve and Drain, Tracy and Hart, William and Imken, Travis and Larson, Kristina and Marsh, Danielle and Muthulingam, Dhack and Snyder, John Steven and Trofimov, Denis and {Elkins-Tanton}, Linda T. and Johnson, Ian and Lord, Peter and Prikl, Zack},
  year = {2019},
  month = sep,
  address = {Vienna, Austria},
  urldate = {2025-10-10}
}

@book{russellDawnMissionMinor2012,
  title = {The {{Dawn Mission}} to {{Minor Planets}} 4 {{Vesta}} and 1 {{Ceres}}},
  author = {Russell, C. T.},
  year = {2012},
  edition = {1st ed},
  publisher = {Springer New York},
  address = {New York, NY},
  collaborator = {Raymond, Carol},
  isbn = {978-1-4614-4902-7 978-1-4614-4903-4},
  langid = {english}
}

@article{chaiReviewOptimizationTechniques2019,
  title = {A Review of Optimization Techniques in Spacecraft Flight Trajectory Design},
  author = {Chai, Runqi and Savvaris, Al and Tsourdos, Antonios and Chai, Senchun and Xia, Yuanqing},
  year = {2019},
  month = aug,
  journal = {Progress in Aerospace Sciences},
  volume = {109},
  pages = {100543},
  issn = {0376-0421},
  doi = {10.1016/j.paerosci.2019.05.003},
  urldate = {2024-07-01}
}

@misc{maoSuccessiveConvexificationSuperlinearly2019,
  title = {Successive {{Convexification}}: {{A Superlinearly Convergent Algorithm}} for {{Non-convex Optimal Control Problems}}},
  shorttitle = {Successive {{Convexification}}},
  author = {Mao, Yuanqi and Szmuk, Michael and Xu, Xiangru and Acikmese, Behcet},
  year = {2019},
  month = feb,
  number = {arXiv:1804.06539},
  eprint = {1804.06539},
  primaryclass = {math},
  publisher = {arXiv},
  doi = {10.48550/arXiv.1804.06539},
  urldate = {2025-06-03},
  archiveprefix = {arXiv},
  langid = {english}
}

@phdthesis{hofmannComputationalGuidanceLowThrust2023,
  title = {Computational {{Guidance}} for {{Low-Thrust Spacecraft}} in {{Deep Space Based}} on {{Convex Optimization}}},
  author = {Hofmann, Christian},
  year = {2023},
  langid = {english},
  school = {Politecnico di Milano}
}

@article{songSolarsailDeepSpace2019,
  title = {Solar-Sail Deep Space Trajectory Optimization Using Successive Convex Programming},
  author = {Song, Yu and Gong, Shengping},
  year = {2019},
  month = jul,
  journal = {Astrophysics and Space Science},
  volume = {364},
  number = {7},
  pages = {106},
  issn = {0004-640X, 1572-946X},
  doi = {10.1007/s10509-019-3597-x},
  urldate = {2025-05-20},
  langid = {english}
}

@article{oguriLosslessControlConvexFormulation2024,
  title = {Lossless {{Control-Convex Formulation}} for {{Solar-Sail Trajectory Optimization}} via {{Sequential Convex Programming}}},
  author = {Oguri, Kenshiro and Lantoine, Gregory},
  year = {2024},
  month = nov,
  journal = {Journal of Guidance, Control, and Dynamics},
  pages = {1--16},
  issn = {0731-5090, 1533-3884},
  doi = {10.2514/1.G008361},
  urldate = {2024-11-26},
  langid = {english}
}

@book{mcinnesSolarSailing1999,
  title = {Solar {{Sailing}}},
  author = {McInnes, Colin Robert},
  year = {1999},
  publisher = {Springer},
  address = {London},
  doi = {10.1007/978-1-4471-3992-8},
  urldate = {2025-09-22},
  copyright = {http://www.springer.com/tdm},
  isbn = {978-1-85233-102-3 978-1-4471-3992-8},
  langid = {english}
}

@article{gongReviewSolarSail2019,
  title = {Review on Solar Sail Technology},
  author = {Gong, Shengping and Macdonald, Malcolm},
  year = {2019},
  month = jun,
  journal = {Astrodynamics},
  volume = {3},
  number = {2},
  pages = {93--125},
  issn = {2522-0098},
  doi = {10.1007/s42064-019-0038-x},
  urldate = {2025-09-22},
  langid = {english}
}

@article{spencerSolarSailingTechnology2019,
  title = {Solar Sailing Technology Challenges},
  author = {Spencer, David A. and Johnson, Les and Long, Alexandra C.},
  year = {2019},
  month = oct,
  journal = {Aerospace Science and Technology},
  volume = {93},
  pages = {105276},
  issn = {1270-9638},
  doi = {10.1016/j.ast.2019.07.009},
  urldate = {2025-09-22}
}

@article{roveyReviewMultimodeSpace2020,
  title = {Review of Multimode Space Propulsion},
  author = {Rovey, Joshua L. and Lyne, Christopher T. and Mundahl, Alex J. and Rasmont, Nicolas and Glascock, Matthew S. and Wainwright, Mitchell J. and Berg, Steven P.},
  year = {2020},
  month = oct,
  journal = {Progress in Aerospace Sciences},
  volume = {118},
  pages = {100627},
  issn = {0376-0421},
  doi = {10.1016/j.paerosci.2020.100627},
  urldate = {2025-10-06}
}

@techreport{manzellaPerformanceEvaluationSPT1401997,
  type = {{{NASA Technical Memorandum}}},
  title = {Performance {{Evaluation}} of the {{SPT-140}}},
  author = {Manzella, David and Sarmiento, Charles and Sankovic, John and Haag, Tom},
  year = {1997},
  month = dec,
  number = {NASA/TM--97-206301},
  institution = {{National Aeronautics and Space Administration, Lewis Research Center}},
  langid = {english}
}

@article{clineIndirectOptimalControl2024,
  title = {Indirect Optimal Control Techniques for Multimode Propulsion Mission Design},
  author = {Cline, Bryan C. and Pascarella, Alex and Woollands, Robyn M. and Rovey, Joshua L.},
  year = {2024},
  month = oct,
  journal = {Acta Astronautica},
  volume = {223},
  pages = {759--776},
  issn = {0094-5765},
  doi = {10.1016/j.actaastro.2024.07.020},
  urldate = {2025-10-05}
}

@article{aryaLowThrustGravityAssistTrajectory2021,
  title = {Low-{{Thrust Gravity-Assist Trajectory Design Using Optimal Multimode Propulsion Models}}},
  author = {Arya, Vishala and Taheri, Ehsan and Junkins, John L.},
  year = {2021},
  month = jul,
  journal = {Journal of Guidance, Control, and Dynamics},
  volume = {44},
  number = {7},
  pages = {1280--1294},
  publisher = {{American Institute of Aeronautics and Astronautics}},
  issn = {0731-5090},
  doi = {10.2514/1.G005750},
  urldate = {2025-10-06}
}

@article{taheriNovelApproachOptimal2020,
  title = {A Novel Approach for Optimal Trajectory Design with Multiple Operation Modes of Propulsion System, Part 1},
  author = {Taheri, Ehsan and Junkins, John L. and Kolmanovsky, Ilya and Girard, Anouck},
  year = {2020},
  month = jul,
  journal = {Acta Astronautica},
  volume = {172},
  pages = {151--165},
  publisher = {Elsevier BV},
  issn = {0094-5765},
  doi = {10.1016/j.actaastro.2020.02.042},
  urldate = {2025-07-09},
  copyright = {https://www.elsevier.com/tdm/userlicense/1.0/},
  langid = {english}
}

@article{malyutaConvexOptimizationTrajectory2022,
  title = {Convex {{Optimization}} for {{Trajectory Generation}}: {{A Tutorial}} on {{Generating Dynamically Feasible Trajectories Reliably}} and {{Efficiently}}},
  shorttitle = {Convex {{Optimization}} for {{Trajectory Generation}}},
  author = {Malyuta, Danylo and Reynolds, Taylor P. and Szmuk, Michael and Lew, Thomas and Bonalli, Riccardo and Pavone, Marco and A{\c c}{\i}kme{\c s}e, Beh{\c c}et},
  year = {2022},
  month = oct,
  journal = {IEEE Control Systems},
  volume = {42},
  number = {5},
  pages = {40--113},
  issn = {1066-033X, 1941-000X},
  doi = {10.1109/MCS.2022.3187542},
  urldate = {2023-10-02},
  langid = {english}
}

@article{rackauckasDifferentialEquationsJlPerformant2017,
  title = {{{DifferentialEquations}}.Jl -- {{A Performant}} and {{Feature-Rich Ecosystem}} for {{Solving Differential Equations}} in {{Julia}}},
  author = {Rackauckas, Christopher and Nie, Qing},
  year = {2017},
  month = may,
  journal = {Journal of Open Research Software},
  volume = {5},
  number = {1},
  pages = {15},
  publisher = {Ubiquity Press},
  issn = {2049-9647},
  doi = {10.5334/jors.151},
  urldate = {2023-07-26},
  langid = {american}
}

@misc{revelsForwardModeAutomaticDifferentiation2016,
  title = {Forward-{{Mode Automatic Differentiation}} in {{Julia}}},
  author = {Revels, Jarrett and Lubin, Miles and Papamarkou, Theodore},
  year = {2016},
  month = jul,
  number = {arXiv:1607.07892},
  eprint = {1607.07892},
  primaryclass = {cs},
  publisher = {arXiv},
  doi = {10.48550/arXiv.1607.07892},
  urldate = {2025-09-22},
  archiveprefix = {arXiv}
}

@article{lubinJuMPRecentImprovements2023,
  title = {{{JuMP}} 1.0: Recent Improvements to a Modeling Language for Mathematical Optimization},
  shorttitle = {{{JuMP}} 1.0},
  author = {Lubin, Miles and Dowson, Oscar and Garcia, Joaquim Dias and Huchette, Joey and Legat, Beno{\^i}t and Vielma, Juan Pablo},
  year = {2023},
  month = sep,
  journal = {Mathematical Programming Computation},
  volume = {15},
  number = {3},
  pages = {581--589},
  issn = {1867-2949, 1867-2957},
  doi = {10.1007/s12532-023-00239-3},
  urldate = {2023-07-26},
  langid = {english}
}

@misc{mosekapsMOSEKOptimizerAPI2025,
  title = {{{MOSEK Optimizer API}} for {{Julia}} 11.0.28},
  author = {{MOSEK ApS}},
  year = {2025}
}

@article{oguriSolarSailingPrimer2022,
  title = {Solar {{Sailing Primer Vector Theory}}: {{Indirect Trajectory Optimization}} with {{Practical Mission Considerations}}},
  shorttitle = {Solar {{Sailing Primer Vector Theory}}},
  author = {Oguri, Kenshiro and Lantoine, Gregory and McMahon, Jay W.},
  year = {2022},
  journal = {Journal of Guidance, Control, and Dynamics},
  volume = {45},
  number = {1},
  pages = {153--161},
  publisher = {{American Institute of Aeronautics and Astronautics}},
  issn = {0731-5090},
  doi = {10.2514/1.G006210},
  urldate = {2023-04-12}
}

@article{lantoineTrajectoryManeuverDesign2024,
  title    = {Trajectory \& Maneuver Design of the {{NEA Scout}} Solar Sail Mission},
  author   = {Lantoine, Gregory and Cox, Andrew and Sweetser, Theodore and Grebow, Dan and Whiffen, Gregory and Garza, David and Petropoulos, Anastassios and Oguri, Kenshiro and Kangas, Julie and Kruizinga, Gerhard and {Castillo-Rogez}, Julie},
  year     = {2024},
  month    = dec,
  journal  = {Acta Astronautica},
  volume   = {225},
  pages    = {77--98},
  issn     = {00945765},
  doi      = {10.1016/j.actaastro.2024.08.039},
  urldate  = {2025-04-29},
  langid   = {english}
}

@misc{hillSparserBetterFaster2025,
  title = {Sparser, {{Better}}, {{Faster}}, {{Stronger}}: {{Sparsity Detection}} for {{Efficient Automatic Differentiation}}},
  shorttitle = {Sparser, {{Better}}, {{Faster}}, {{Stronger}}},
  author = {Hill, Adrian and Dalle, Guillaume},
  year = {2025},
  month = jun,
  number = {arXiv:2501.17737},
  eprint = {2501.17737},
  primaryclass = {cs},
  publisher = {arXiv},
  doi = {10.48550/arXiv.2501.17737},
  urldate = {2025-10-15},
  archiveprefix = {arXiv}
}

@article{walkerSetModifiedEquinoctial1985,
  title = {A Set of Modified Equinoctial Orbit Elements},
  author = {Walker, M. J. H. and Ireland, B. and Owens, Joyce},
  year = {1985},
  month = aug,
  journal = {Celestial Mechanics},
  volume = {36},
  number = {4},
  pages = {409--419},
  issn = {1572-9478},
  doi = {10.1007/BF01227493},
  urldate = {2023-06-19},
  langid = {english}
}
\end{document}